\documentclass[a4paper,12pt]{article}
\usepackage[english]{babel}
\usepackage[T1]{fontenc}
\usepackage{graphics}
\usepackage{booktabs}
\usepackage{amsmath}
\usepackage{inputenc}
\usepackage{amsthm}
\usepackage{amssymb}
\usepackage{setspace}
\usepackage{array}
\usepackage{graphicx}
\usepackage{subfig}
\usepackage{varioref}
\usepackage[blocks]{authblk}
\usepackage[font=footnotesize,format=hang]{caption}
\usepackage[a4paper,top=3cm,bottom=3cm,left=2.5cm,right=2.5cm,%
bindingoffset=5mm]{geometry}
\usepackage{url}
\addcontentsline{toc}{section}{Acknowledgement}

\theoremstyle{plain}

\theoremstyle{remark}

\title{Exploring the relationship between the magnitudes of seismic events}

\date{}
\usepackage{authblk}

\author[1]{Ilaria Spassiani}
\author[1]{Giovanni Sebastiani}
\affil[1]{Department of Mathematics ``Guido Castelnuovo'', Sapienza University of Rome, Rome, ITALY.}
\affil[2]{Istituto per le Applicazioni del Calcolo ``M. Picone'',
Consiglio Nazionale delle Ricerche, Rome, ITALY.}

\begin{document}
\maketitle

\begin{abstract}
The distribution of the magnitudes of seismic events is generally assumed to be independent on past seismicity. However, by considering events in causal relation, for example mother-daughter, it seems natural to assume that the magnitude of a daughter event is conditionally dependent on the one of the corresponding mother event. In order to find experimental evidence supporting this hypothesis, we analyze different catalogs, both real and simulated, in two different ways. From each catalog, we obtain the law of triggered events' magnitude by kernel density. The results obtained show that the distribution density of triggered events' magnitude varies with the magnitude of their corresponding mother events. As the intuition suggests, an increase of mother events' magnitude induces an increase of the probability of having ``high'' values of triggered events' magnitude. In addition, we see a statistically significant increasing linear dependence of the magnitude means.
\end{abstract}

\section{Introduction}
The Epidemic Type Aftershock Sequences (ETAS) is a well known model in seismology \cite{ogata:primo,ogata:secondo,ogata:terzo,ogata:quarto}. It is a branching process in which, at each generation, each event may produce its own offsprings independently of the others.
The magnitudes of events are mutually statistically independent and are distributed according to the \textit{Gutenberg-Richter law} \cite{gr:primo}. The distribution density function of an event's magnitude $m$ is given by $p(m)=b\ln10\cdot10^{-b(m-m_0)}=\beta e^{-\beta(m-m_0)}$, where $\beta=b\ln10$ is a positive constant and the cutoff parameter $m_0$ is known as the \textit{completeness magnitude}. This parameter is estimated from data in such a way that magnitudes exceeding it, are statistically in very good agreement with the Gutenberg-Richter law. This law is assumed to be valid both for background events and for the triggered ones and is independent on the magnitude of the corresponding ancestors \cite{zhuang:primo}. Recall that the background seismicity is the component not triggered by precursory events and is usually connected to the regional tectonic strain rate; on the other hand, the triggered seismicity is the one associated with stress perturbations due to previous shocks \cite{marzocchi:primo}.

In this work we will use the terms ``triggering event'' to indicate any mother event that produces its own progeny \cite{sornettesaichev:primo}. Both a triggered and a background event may be a triggering shock.
We want to investigate the distribution of the magnitude of triggered events, in order to assess its variation with the magnitude of the corresponding triggering ones. In fact, even if the correlation between subsequent events is very difficult to be detected and therefore is usually assumed to be absent \cite{helm:primo,corral:primo}, in some recent works it was found statistically different from zero \cite{lippi:primo,lippi:secondo,lippi:terzo,sarlis:primo,sarlis:secondo}. We expect that, the distribution density of triggered events' magnitude increases (decreases) with the increase (decrease) of triggering events' magnitude. Moreover, if this is true, the expected value of the triggered events' magnitude should also be increasing in the same way.

\section{Materials and methods}
In order to obtain experimental evidence to support the above-mentioned hypothesis, we perform two different kinds of analysis of three Italian seismic catalogs, a Californian catalog and some simulated ones. The real data catalogs used here are the following.
\begin{enumerate}
\item The first catalog includes events occurred from 16 April 2005 till 25 January 2012 in the whole Italy. The estimated value for the completeness magnitude is 2.5.
\item The second catalog includes events occurred in the portion of Abruzzo region (Italy) corresponding to the square from latitude +41.866 to +42.866 and from longitude +12.8944 to +13.8944. This subregion includes L'Aquila. The temporal interval is the same as for catalog one. The estimated value for the completeness magnitude is 1.8. This catalog includes the strong shock of magnitude 6.1 occurred in L'Aquila on April the $6^{th}$, 2009.
\item The third catalog differs from the previous one only for the temporal interval, which now goes from 16 April 2005 till 05 April 2009. The value of completeness magnitude is estimated equal to 1.5. This catalog doesn't include a strong shock.
\item The fourth catalog includes events occurred from 01 January 1984 till 31 December 1991, in the portion of the Southern California corresponding to the square from latitude +33.75 to +34.75 and from longitude -117.5 to -116.5. The estimated completeness magnitude is 2. This catalog is a portion of the waveform earthquake catalog relocated by Hauksson \textit{et al.} in 2011 \cite{hauksson:primo,hauksson:secondo}.
\end{enumerate}
In all the previous cases the maximum depth considered is set equal to 40 km.

In order to find empirical evidence for the hypothesis that the distribution of triggered events' magnitude depends on their own triggering events' one, it seems appropriate not to include in the catalog events that are spatially ``too'' distant to each other. This is the reason for including in the study catalogs two, three and four. More precisely, we consider the third catalog to investigate the influence of a strong shock and the fourth one to verify the validity of our hypothesis also for a catalog relative to a region far away from the others and with a different seismicity. Instead, in the first catalog there is the presence of events that are very distant to each other, as can be seen from the mean values of the distances between the events considered by us in causal relation, shown below. We expect that in this case the above-mentioned dependence becomes far less evident or absent.

The values of the completeness magnitude have been computed with the ZMAP software, using Shi and Bolt uncertainty \cite{shi:primo}.

The simulated catalogs have been considered, instead, to demonstrate that our hypothesis is true also for pure simulated catalogs, not affected by any kind of ``real effect'' that may influence the analysis. The results reported here are referred to two simulated catalogs obtained as follows. The first one is simulated with the FORTRAN program [\textbf{etasim.f}] written by Ogata \cite{ogata:secondo,ogata:quinto,ogata:sesto}, just modified in the fact that the number of events is random between the two input starting and ending times. We have chosen here the option of simulating the magnitudes with the Gutenberg-Richter law, instead of taking them from a given catalog. Then, we expect that there is no evidence of our hypothesis of conditioning. On the other hand, we simulate the second synthetic catalog with a program written by us very similar to the Ogata's one, but adapted to our hypothesis of conditioning. More precisely, we simulate the magnitudes of background events again with the Gutenberg-Richter law, while the magnitudes of the triggered events with a new conditional probability density function with respect to the triggering events' magnitudes. Since we expect that, when the triggering events' magnitude $m'$ increases, the probability of having events with ``high'' magnitudes must increase and, at the same time, the one of events with ``low'' magnitudes must decrease, we propose the following probability density function:
\begin{equation}
\label{eqn:pcond}
p(m|m')=\beta e^{-\beta(m-m_0)}\biggl[1+C_1\bigl(1-2e^{-(\beta-a)(m'-m_0)}\bigr)\bigl(1-2e^{-\beta(m-m_0)}\bigr)\biggr],
\end{equation}
where $\beta$ and $a$ are the parameters of the Gutenberg-Richter law and the \textit{productivity law} $\rho(m)=\kappa e^{a(m-m_0)}$, respectively, and $1>C_1\ge0$. We notice that this model reduces to the Gutenberg-Richter one for $C_1=0$. For the theoretical motivation and properties about this formula, see \cite{mio:primo}.

In both the two simulated catalogs, we consider the same minimum magnitude value equals to 1.5 and a null learning (precursory) period.

For each catalog, in the two types of analysis we consider four magnitudes subintervals contained in the magnitude range, from the completeness magnitude to the maximum value. The amplitude of each subinterval is opportunely chosen, for each catalog, in such a way in order to have a comparable number of triggered events in all the subintervals considered.

We then proceed differently for the two types of analysis.
In the first approach, we group the events whose magnitude belongs to each of the above subintervals. We now consider the events of any given subinterval. For each of them, we look at all the magnitudes of the events occurred within a certain time interval of amplitude ${\delta}^*$ after it. We than group these magnitudes corresponding to all the ``starting'' events. Based on this set of magnitudes, indicated in what follows with $G_m$, we estimate a probability density function using kernel density estimation \cite{rosen:primo}. This is a non-parametric method very used in statistical analysis. Precisely, we consider the magnitudes $(m_1, m_2,\dots, m_n)$ and the frequencies $f=(f_1, f_2,\dots, f_n)$ of $G_m$. We then consider the set $m$ of 1000 magnitudes equispaced from the completeness magnitude to the maximum one and we compute the \textit{kernel density estimator} of the empirical magnitude distribution $M$ as
\begin{equation}
\label{eqn:1}
\hat{M}_{\gamma}(m)=\frac{1}{ F}\sum_{i=1}^n f_i K\biggl(\frac{m-m_i}{\gamma}\biggr),\quad\text{with}\quad F=\sum_{i=1}^nK\biggl(\frac{m-m_i}{\gamma}\biggr),
\end{equation}
where $K(\cdot)$ is known as \textit{kernel} and the positive parameter $\gamma$ is the \textit{bandwidth} \cite{parzen:primo}. The above formula is obtained by adapting to our case the \textit{Nadaraya-Watson} kernel for kernel regression \cite{scott:primo}.

Let's remind that a kernel is a non-negative, real-valued function such that
\[
\int_{-\infty}^{\infty}K(x)dx=1,\quad\text{and}\quad K(x)=K(-x)\quad\forall x\in\mathbb{R}.
\]
As very often done, here we use the Gaussian kernel
\[
K(x)=\frac{1}{\sqrt{2\pi}}e^{-\frac{x^2}{2}}.
\]
The value of the bandwidth is chosen using the leave-one-out cross-validation method, opportunely implemented by us \cite{scott:primo}. Precisely, we consider the value that minimizes the quantity $\sum_{i=1}^n |\hat{f}_i-f_i|$, where
\[
\hat{f}_{i}=\frac{1}{ \bar{F}_i}\sum_{j\neq i} f_j K\biggl(\frac{m_i-m_j}{\gamma}\biggr),\quad\text{with}\quad\bar{F}_i=\sum_{j\neq i}K\biggl(\frac{m_i-m_j}{\gamma}\biggr).
\]
We observe that, differently from the formula~\eqref{eqn:1}, the value $f_i$ doesn't contribute to $\hat{f}_{i}$.

The time window ${\delta}^*$ is chosen here in such a way that two seismic events separated by a time larger than ${\delta}^*$ are not in causal relation. In order to determine this value, for each catalog we divide the whole time interval in daily subintervals. We then count the number of events that occur in each subinterval. Starting from this temporal sequence, denoted by $X_t$, we then compute an estimate $\widehat{R}(\delta)$ of the autocorrelation function at different integer values of the time lag $\delta$:
\[
\widehat{R}(\delta)=\frac{1}{(n-\delta)\widehat{V}}\sum_{t=1}^{n-\delta}(X_{t}-\widehat{\mu})(X_{t+\delta}-\widehat{\mu}),
\]
where $n$ is the dimension of the sample, $\delta=0,1,2,\dots$ and $\widehat{\mu}=\frac{1}{n}\sum_{t=1}^nX_t$ and $\widehat{V}=\frac{1}{n-1}\sum_{t=1}^n(X_t-\widehat{\mu})^2$ are the sample mean and variance, respectively \cite{priest:primo}.

Then, we model $\widehat{R}(\delta)$ by a power law model containing two parameters. These parameters are estimated by least squares. Finally, we find the value ${\delta}^*$ such that, for lag values larger than ${\delta}^*$, the model is less than $5*10^{-2}$. We notice that, in the cases examined, this choice produces \textit{p-values} always smaller than 0.01.

Due to the strong shock on April the $6^{th}$, 2009, the second catalog shows a clear non-stationary pattern. Then, for this catalog we transform the original dataset by considering the well-known random time change:
\begin{equation}
\label{eqn:2}
\int_0^t \lambda(x)dx,
\end{equation}
where $\lambda(\cdot)$ is the ETAS rate for seismic events. The parameters of the above formula have been computed with the FORTRAN program [\textbf{etas.f}] written by Ogata, by which statistical inference on the parameters of the ETAS model is performed \cite{ogata:sesto}. By this transformation, the process becomes stationary.

In the second type of analysis we proceed as follows. At first we apply the same Ogata's program as above to estimate the parameters. For each event of the catalog, we then find the mother shock that most likely triggered it, by a variation of the Ogata's criterion. More precisely, we consider as mother the preceding event that gives the highest contribute to the ETAS rate. After that, we consider the magnitudes subintervals strategy as in the first type of analysis. For each of them, we group the triggered events whose triggering shock's magnitude belongs to the considered subinterval. As in the first kind of analysis, for each subinterval we estimate the probability density function relative to triggered events' magnitude by using the Gaussian kernel density estimation method described above. The value of the bandwidth is determined as before.

We do not apply this analysis to the first catalog. In fact, it is not meaningful to use the pure temporal ETAS model for such a large region like the one in this catalog. We do not show here the results obtained for this catalog with the first analysis, which are qualitatively similar to those in the left plot at the top of Fig.~\ref{fig:subfig6}. In fact, also in this case if the plot the estimated densities of the triggered events' corresponding to four magnitude subintervals opportunely chosen, we can see no apparent differences among the densities. As said before, this can be explained by the fact that there are many pairs of events that are close to each other along time, but spatially very separated. The elements of these pairs are erroneously put in relation in the analysis. In this case, the mean distance between the events of the pairs in causal relation is 140 Km.

We implemented both methods of analysis in the MATLAB language.

\section{Results}
We present here some results obtained by the two types of the above-mentioned analysis for the catalogs considered.

In Fig.~\ref{fig:subfig2}, there are the estimated densities of the triggered events' magnitude in the real catalogs, obtained by the first and the second types of analysis (left and right plots, respectively). At the top, we show the results relative to the second catalog (L'Aquila till 2012). In this case, the spatial extension of the region analysed is far smaller than the one of the whole Italian catalog. We recall that, in this case, the first analysis has been applied to the dataset transformed by the random time change~\eqref{eqn:2}, due to the non-stationary pattern of the process caused by the presence of the strong shock on April the $6^{th}$, 2009. The mean of the distances between the events of the pairs is about 7 Km and 13 Km for the first and the second types of analysis, respectively. From the first analysis (left plot), we notice that the increase of the referential magnitude corresponds to a qualitative variation of the density in agreement with our hypothesis. In fact, there is the increase (decrease) of the density for high (low) values of the magnitude. The results for the second analysis (right plot) show the same qualitative variations. The learning period, chosen to estimate the parameters, ends at the time of the last event occurred on April the $5^{th}$, 2009. We get $(\mu,\kappa,c,a,p)=(0.304,0.06,0.104,1.57,1.39)$, where $c$ and $p$ are the parameters of the \textit{Omori-Utsu law} $\phi(t)=(t+c)^{-p}$.

In the middle of Fig.~\ref{fig:subfig2}, there are the results concerning the third catalog. In this case the temporal period is shorter than for the second one and ends the day before the strong shock on April the $6^{th}$, 2009. The means of the distances between the events of the pairs are 23 Km and 17 Km for the first and the second types of analysis, respectively. Due to the absence of a very strong shock in this catalog, the parameters used here are the averages over all the sets of parameters obtained by setting the learning period to 7\%, 8\%, $\dots$, 20\%. We get $(\mu,\kappa,c,a,p)=(0.5893,0.0219,0.0151,1.6521,1.1186)$. The parameters values corresponding to the different precursory periods show small variations from the above means. The results are qualitatively very close to those of the two plots at the top. This shows that our hypothesis of dependence is not related to the presence of a strong shock. Furthermore, recalling also that the completeness magnitude is smaller for the third catalog, we can conclude, according to \cite{lippi:quarto}, that our hypothesis is not even connected to the incompleteness of the catalog, as instead proposed by \cite{corral:secondo}.

At the bottom of Fig.~\ref{fig:subfig2}, we can see the results relative to the fourth catalog, that is the Californian one. Here, the mean of the distances is 13 Km for both the two types of analysis. Both from the left and the right plots, respectively obtained with the first and the second types of analysis, we get results in agreement with the above behaviors. It follows that, even if we analyze the events of a region in another continent, the hypothesis is still supported by the results of the two types of analysis. The parameters are here again obtained by averaging over the sets estimated for a learning period fixed at 7\%, 8\%, $\dots$, 20\%. We get $(\mu,\kappa,c,a,p)=(0.3729,0.0116,0.0002,0.8579,0.8879)$. Again the values obtained for the different learning periods are close to these mean values.

We are going now to analyze the results obtained for the two synthetic catalogs.

Fig.~\ref{fig:subfig6} contains the estimated densities of triggered events' magnitude, obtained by the first and the second types of analysis (left and right plots, respectively), relative to the simulated catalogs. At the top, we plot the results concerning the catalog simulated with the classical Ogata's model. In this case, the magnitudes are computed randomly with the Gutenberg-Richter law. According to this law, triggered events' magnitudes aren't correlated with their respective mothers' magnitudes. This is reflected in the absence of variations of the densities in the four magnitude subintervals considered, both in the left and the right plots (first and second types of analysis, respectively). The parameters, estimated by setting the precursory at about 10\%, are $(\mu,\kappa,c,a,p)=(0.62,0.02,0.013,1.72,1.11)$.

Instead, a result that strongly supports our hypothesis of correlation is the one shown in the two plots at the bottom of Fig.~\ref{fig:subfig6}. Here, we show the estimated densities of triggered events' magnitude for the catalog simulated with our model. That is, the catalog in which the magnitudes are computed with the conditional probability density function~\eqref{eqn:pcond}. Once obtained this catalog, we estimated the parameters with the classical Ogata's FORTRAN program [\textbf{etas.f}], fixing again the learning period at about 10\%. We get $(\mu,\kappa,c,a,p)=(0.58,0.022,0.017,0.83,1.12)$. The behavior in these figures is exactly the same as the one we obtain for the real data. This strongly supports our hypothesis.

Finally, Fig.~\ref{fig:subfig4} contains the plots of the averages of triggered events' magnitudes versus triggering events' magnitudes for both types of analysis (respectively left and right plots). In the two plots at the top, we consider the catalog simulated with the Ogata's model and the one simulated with our model (red and black lines, respectively). For each of the two catalogs considered, the four triggered magnitude averages are normalized by the means of the average of the four values. The results of the linear regression analysis and the error bars are also shown. The lengths of the latter are given by the normalized mean standard errors. Regarding both the left and the right plots, one can see that there is almost no percentage variation of the triggered events' magnitude in the catalog simulated with the Ogata's model. Instead, a clear increasing trend is evident for the other simulated catalog considered.

The two plots at the bottom contains the results concerning the second and the third catalogs, relative to L'Aquila till 2012 (blue line) and L'Aquila till 5 April 2009 (magenta line), respectively. The means have an increasing trend in both of these cases. The results for the Italian and the Californian catalogs are not shown here, but we can say that the Californian catalog exhibits the same behavior of the plots at the bottom here. On the other hand, the means obtained with the first analysis for the whole Italian catalog are similar to those obtained for the catalog simulated with the Ogata's model, as expected for the reasons explained before.

The results obtained are statistically significant, as one can see from the following list of correlation coefficient $R$ and p-values $p$. First analysis: catalog one, $R\simeq0.88$ and $p\simeq0.11$; catalog two, $R\simeq0.99$ and $p\simeq0.004$; catalog three, $R\simeq0.96$ and $p\simeq0.03$; catalog four, $R\simeq0.99$ and $p\simeq0.0005$; catalog simulated with the Ogata's model, $R\simeq0.61$ and $p\simeq0.38$; catalog simulated with our model, $R\simeq0.99$ and $p\simeq0.002$. Second analysis: catalog two, $R\simeq0.94$ and $p\simeq0.05$; catalog three, $R\simeq0.94$ and $p\simeq0.05$; catalog four, $R\simeq0.95$ and $p\simeq0.04$; catalog simulated with the Ogata's model, $R\simeq-0.92$ and $p\simeq0.07$; catalog simulated with our model, $R\simeq0.98$ and $p\simeq0.01$.

\section{Conclusion}
In order to find evidence to support the dependence of the triggered events' magnitude distribution on the triggering events' magnitude, we have applied two types of analysis to some catalogs, both real and simulated. In the well-known ETAS model the distribution of the triggered events' magnitude is the same as that of the triggering events' one and is independent on past seismicity. Instead, our results support the intuitive and more realistic hypothesis of above. We notice that the probability density function for triggered events' magnitude seems to vary when triggering events' magnitude increases. In particular, we observe the increase (decrease) of the density for ``high'' (``low'') values of the magnitude density. This is true for all the catalogs with the exception of the whole Italian catalog and the one simulated by the classical Ogata's model. For the first catalog, the absence of the variation is due to the fact that it contains many pairs of events temporally close to each other, but spatially very separated. For the simulated catalog just mentioned, the variation is absent since it is obtained using the standard Gutenberg-Richter law for the magnitudes. The fact that the catalog simulated by our magnitude model shows a clear evidence of the variations, qualitatively similar to those of the real catalogs, gives strong support to the validity of our hypothesis. Regarding the triggered events' magnitude averages, we can see that they have always an increasing trend, again with the exception of the whole Italian catalog and the one simulated by the Ogata's model. We interpret the results for the remaining catalogs as done before. Concluding, the law of triggered events' magnitude should be a conditional probability density function that changes in shape with the triggering events' magnitude. More precisely, when the latter increases, it may have some relative maximum for higher values of the density. It should also have an increasing expected value.

\section*{Acknowledgments}
We would like to thank Dr. A. Govoni, Dr. W. Marzocchi and specially Dr. A. Lombardi of the Istituto Nazionale di Geofisica e Vulcanologia for their very useful comments and suggestions. 

The Italian data are available through the online data-base ISIDe, Italian Seismological Instrumental and Parametric Data-base (\url{http://iside.rm.ingv.it/}).

The Californian data are available through the online data-base SCEDC, Southern California Earthquake Data Center 

(\url{http://www.data.scec.org/research-tools/alt-2011-dd-hauksson-yang-shearer.html}).

\clearpage
\addcontentsline{toc}{section}{\refname}
\nocite{*}
\bibliographystyle{plain}
\bibliography{biblio}

\newpage

\section{Figure captions}
\begin{figure}
   \centering
\mbox{\includegraphics[width=20pc]{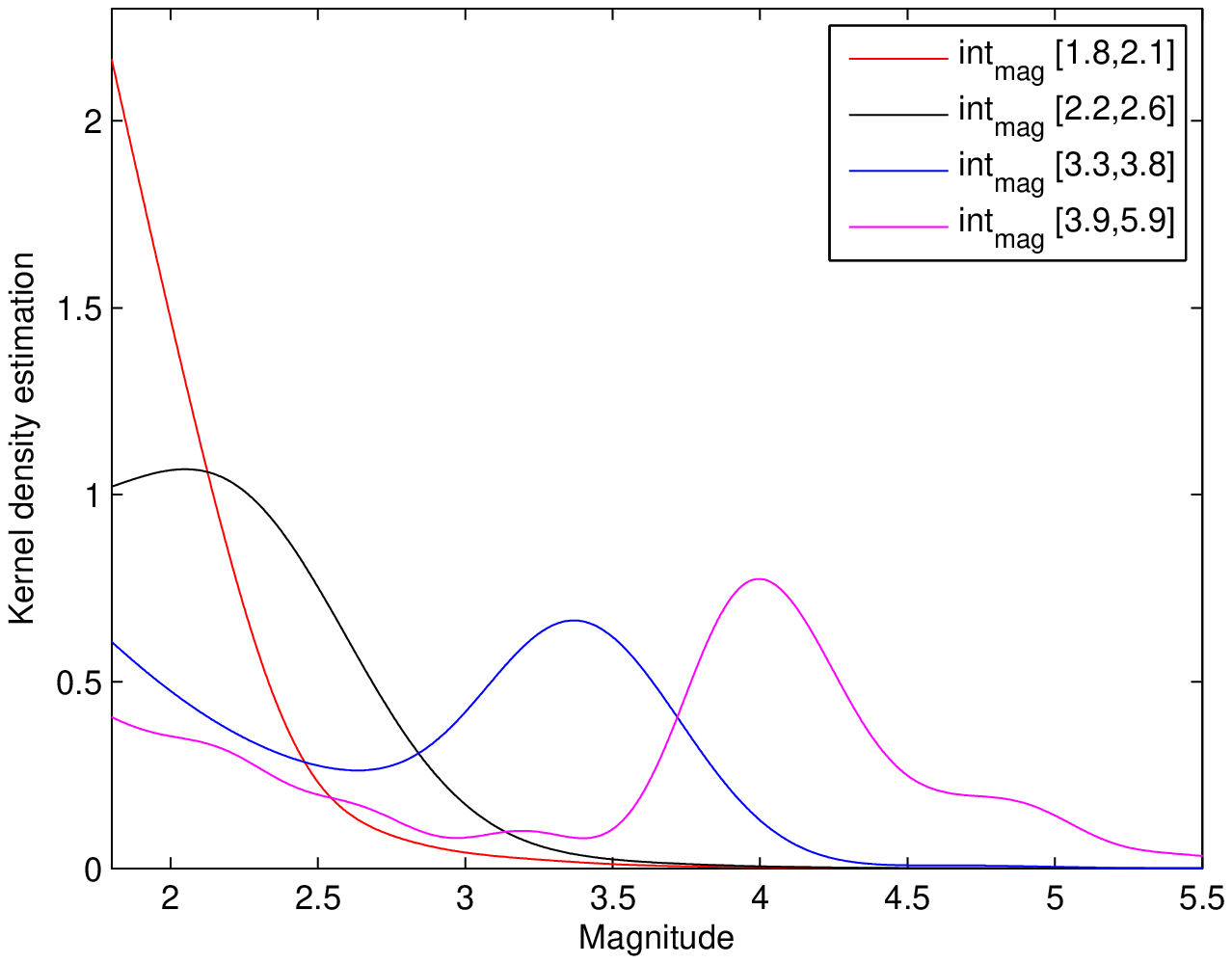}
\includegraphics[width=20pc]{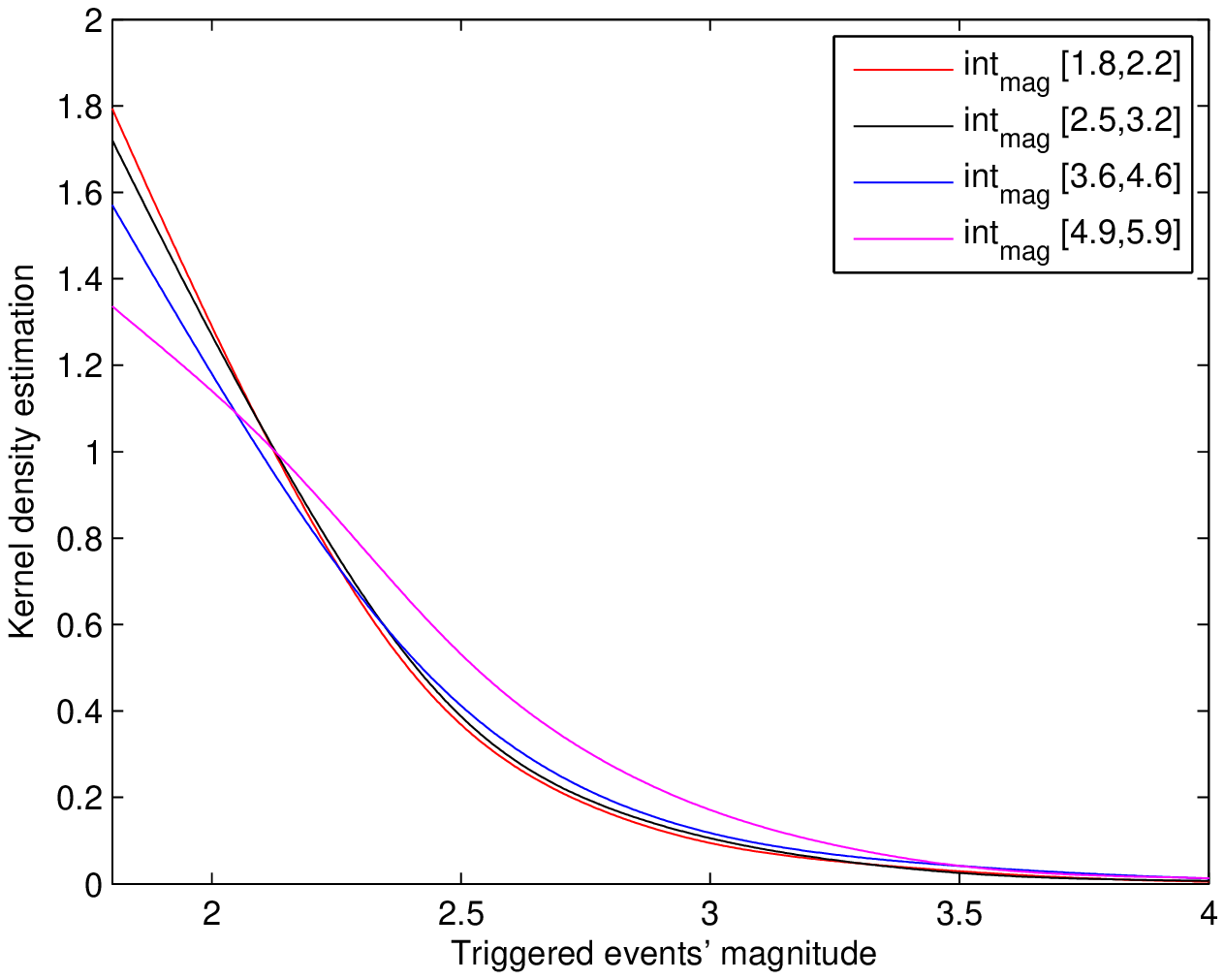}}
\mbox{\includegraphics[width=20pc]{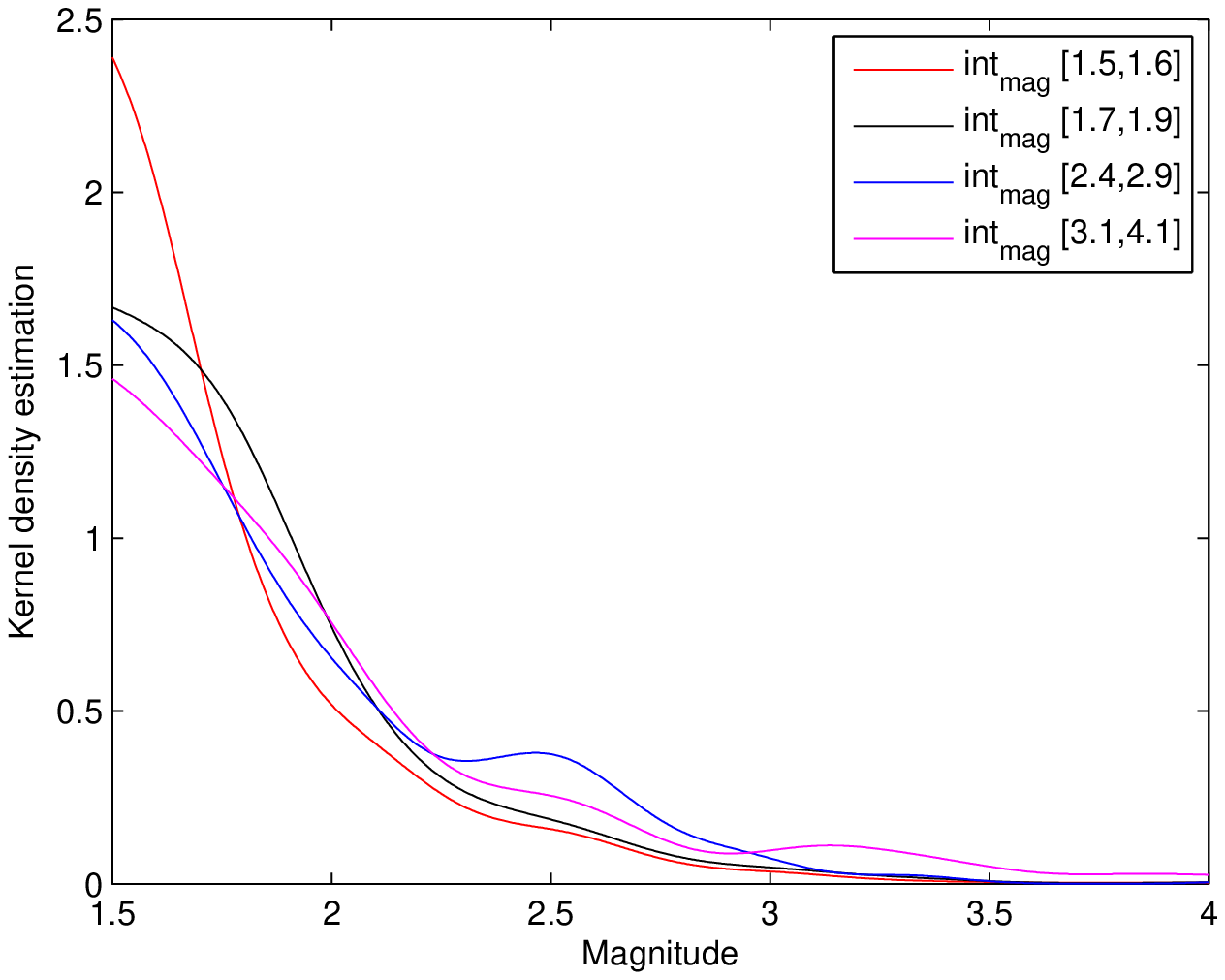}
\includegraphics[width=20pc]{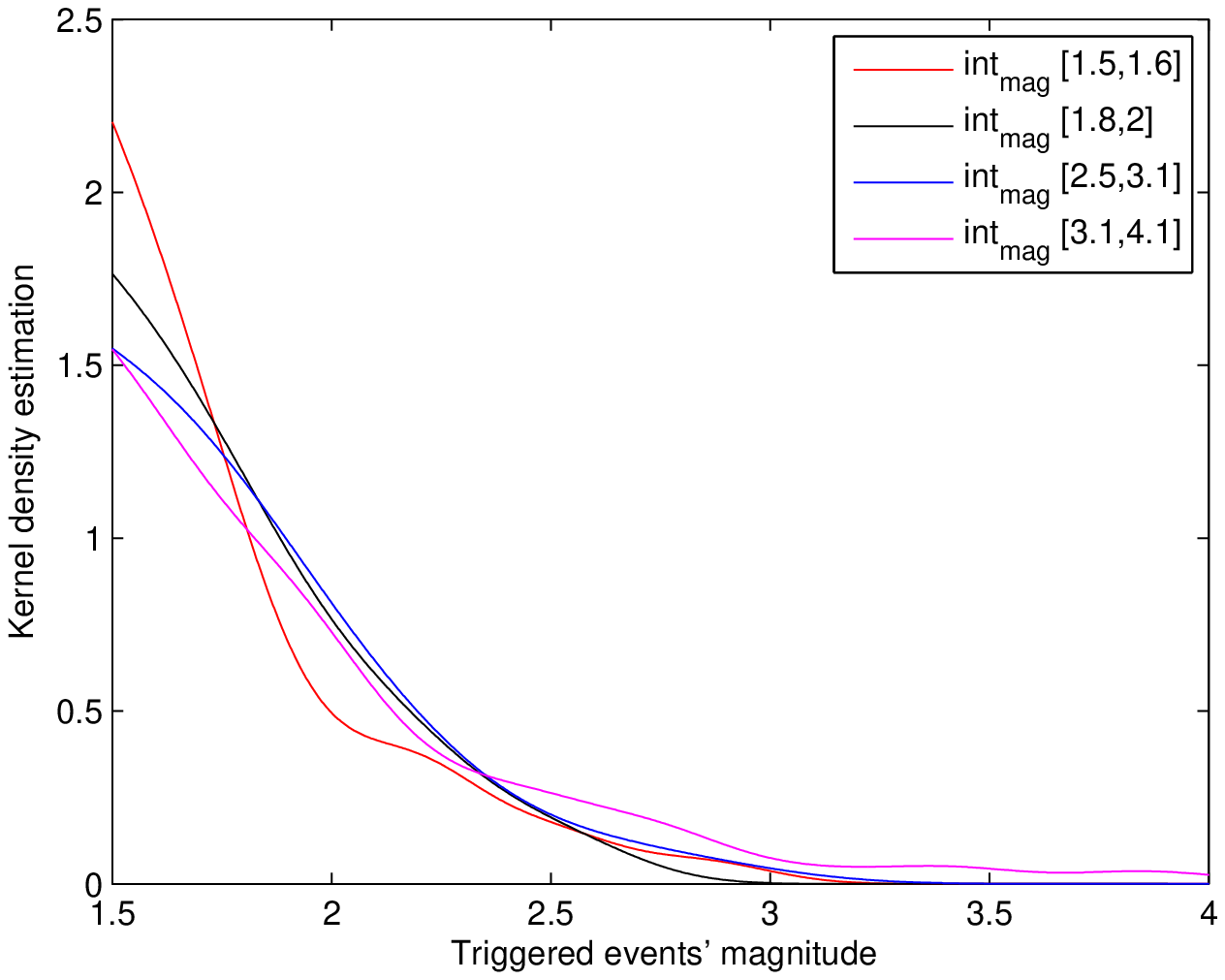}}
\mbox{\includegraphics[width=20pc]{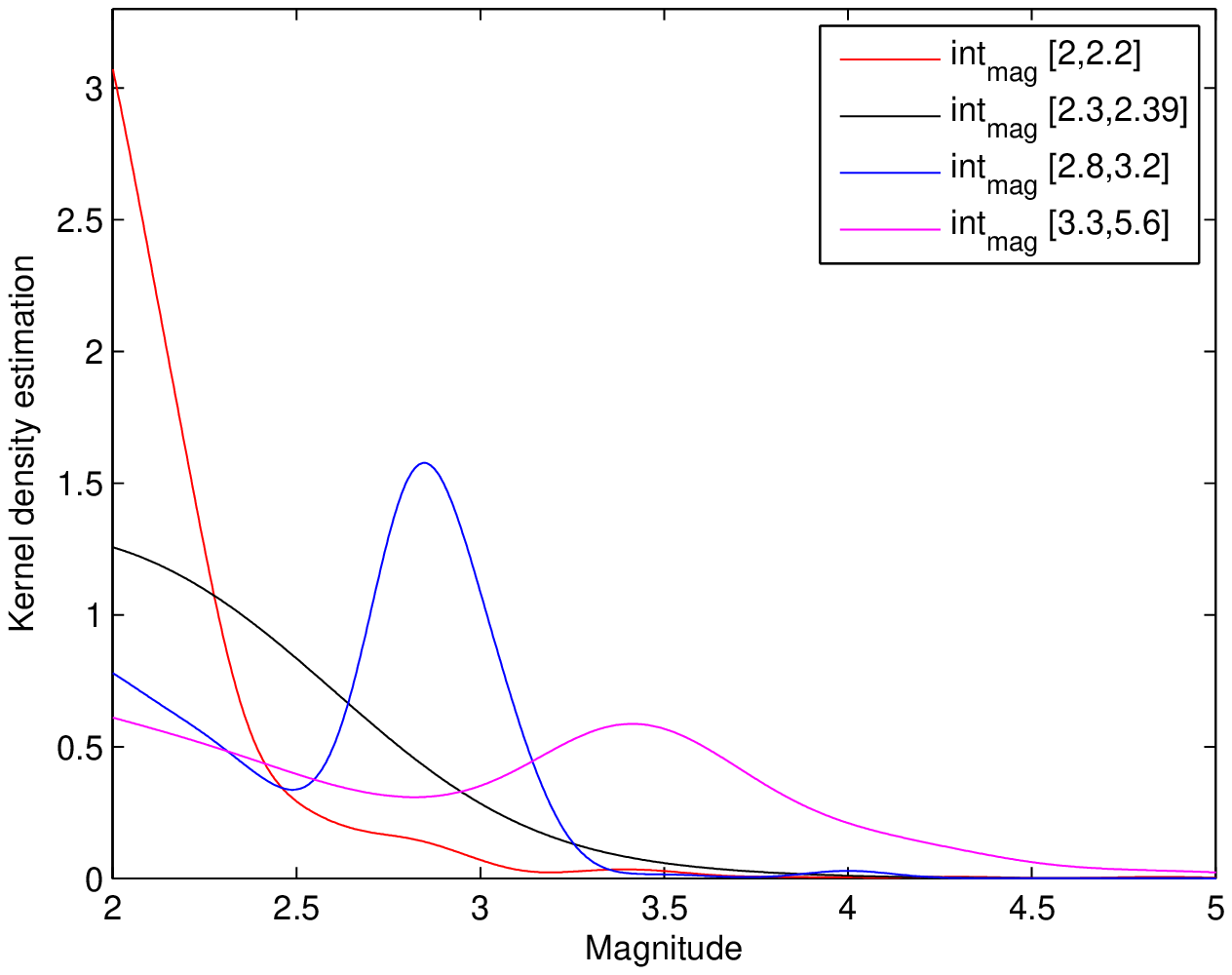}
\includegraphics[width=20pc]{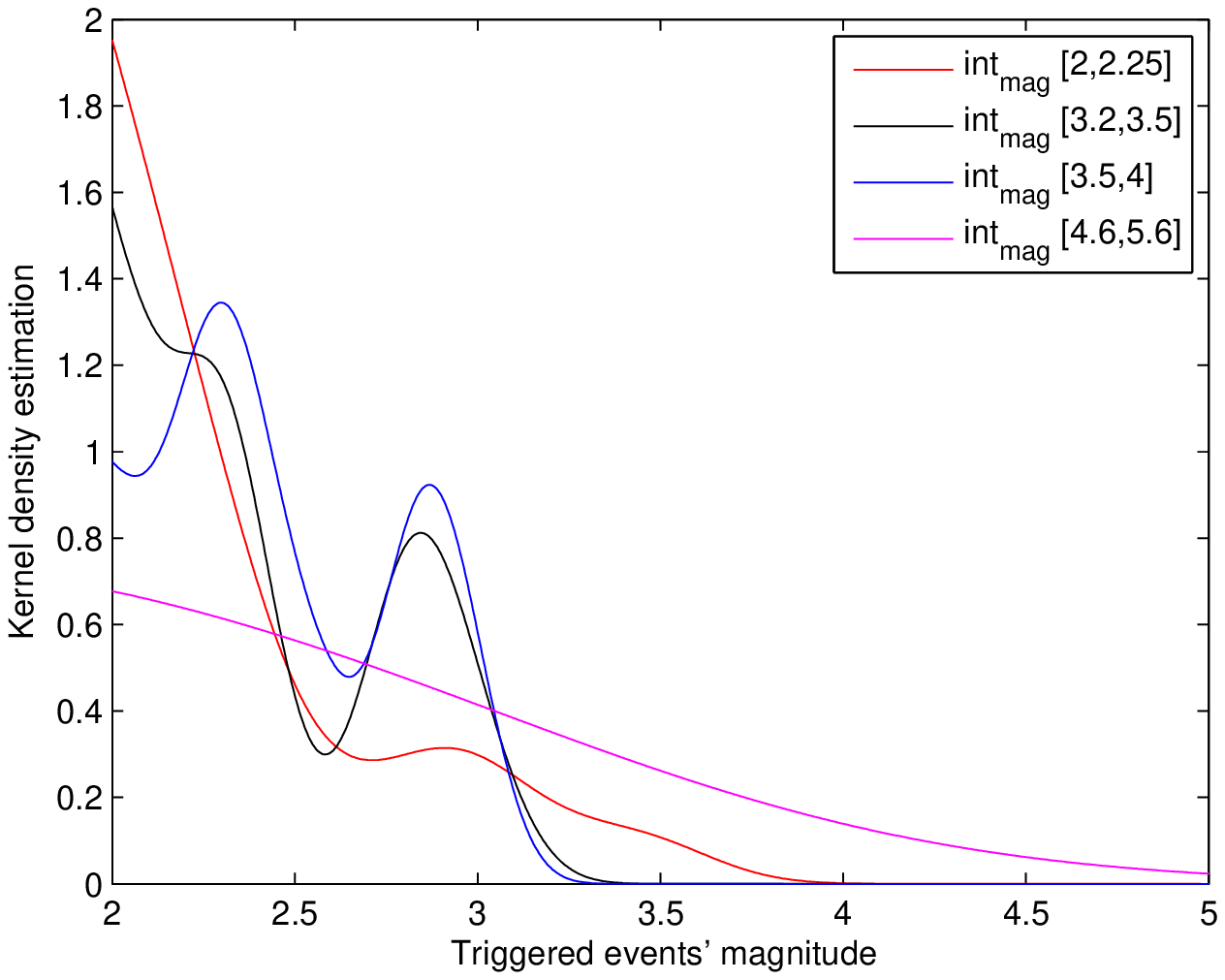}}
   \caption{}
\label{fig:subfig2}
\end{figure}

\begin{figure}
   \centering
\mbox{\includegraphics[width=20pc]{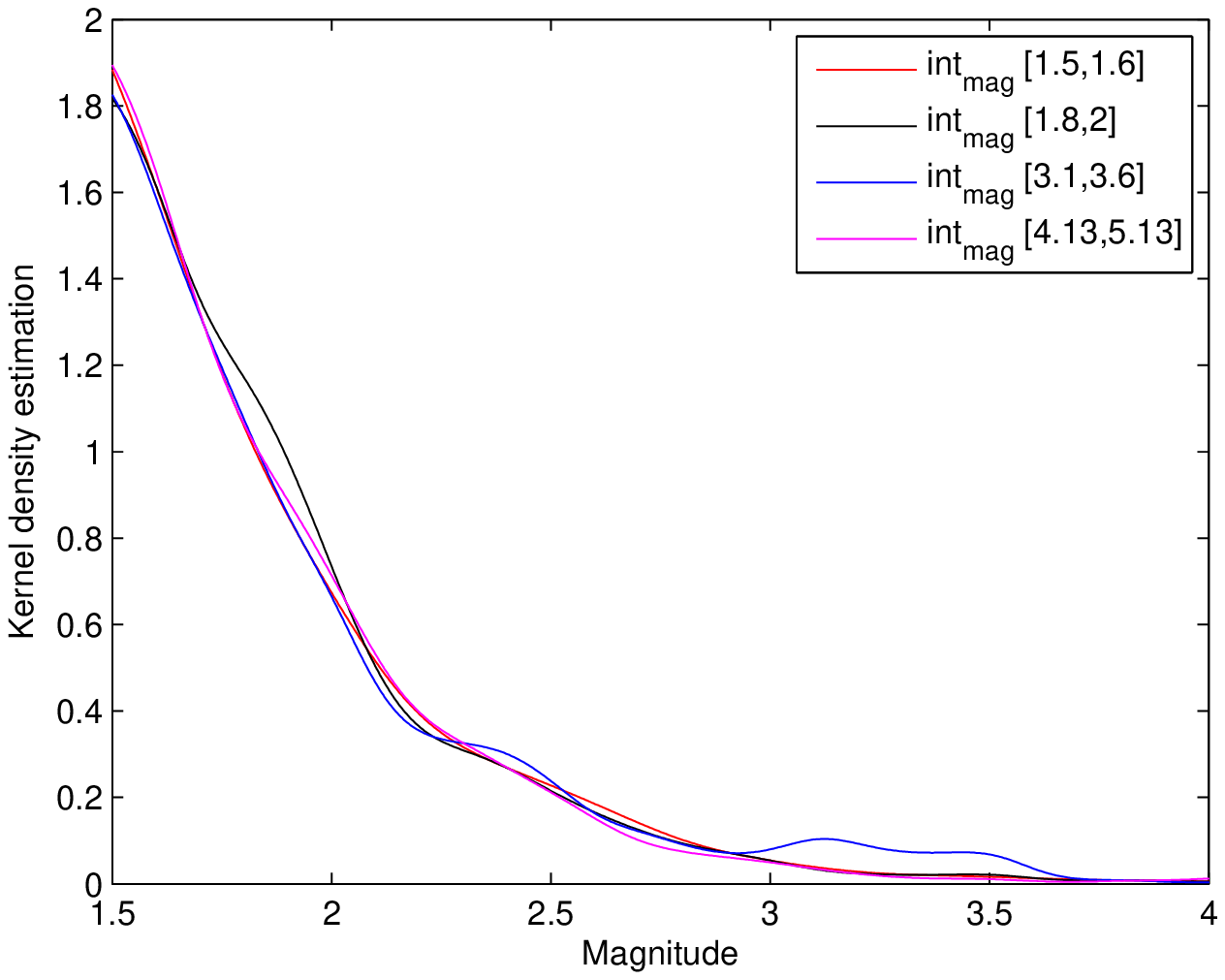}
\includegraphics[width=20pc]{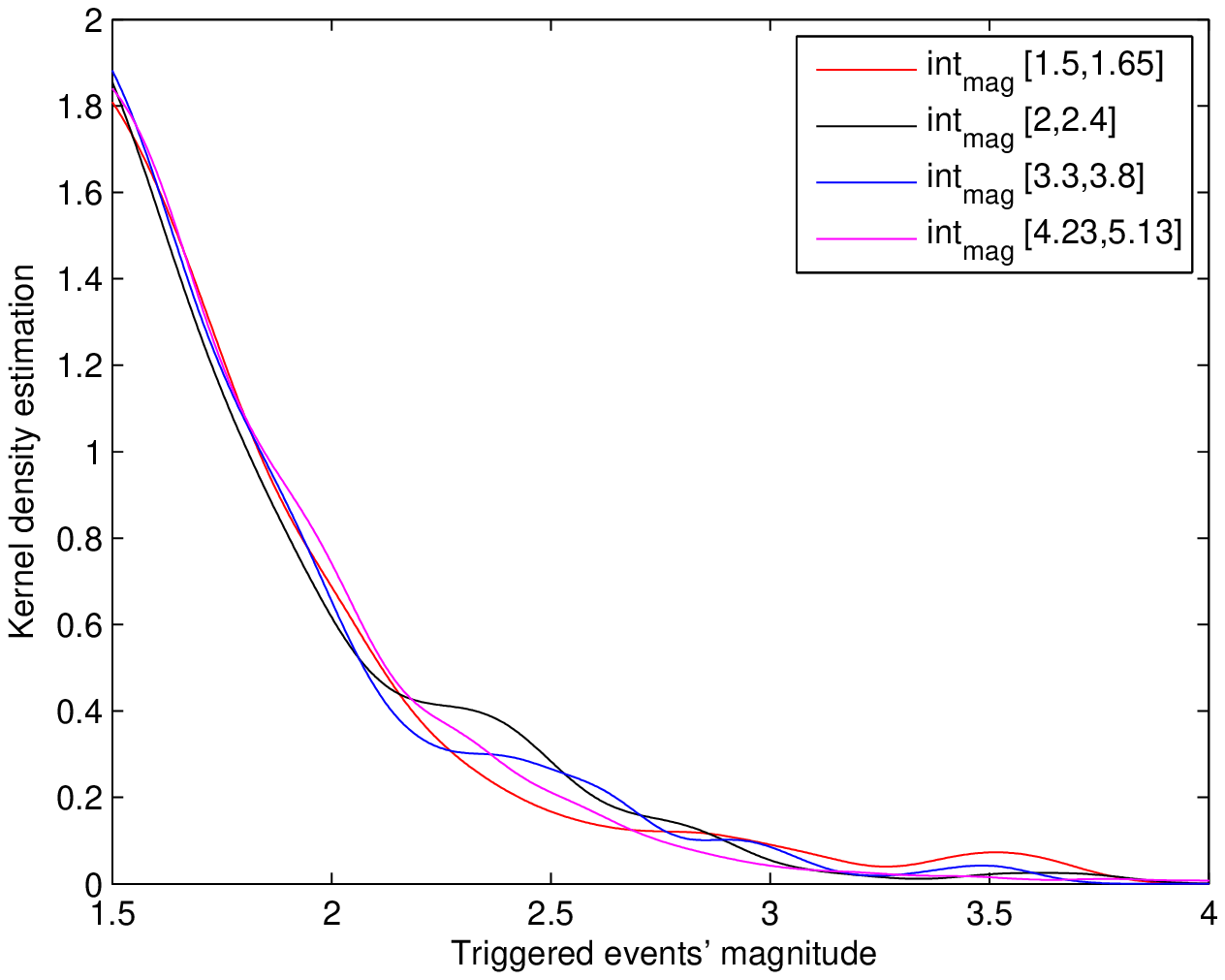}}
\mbox{\includegraphics[width=20pc]{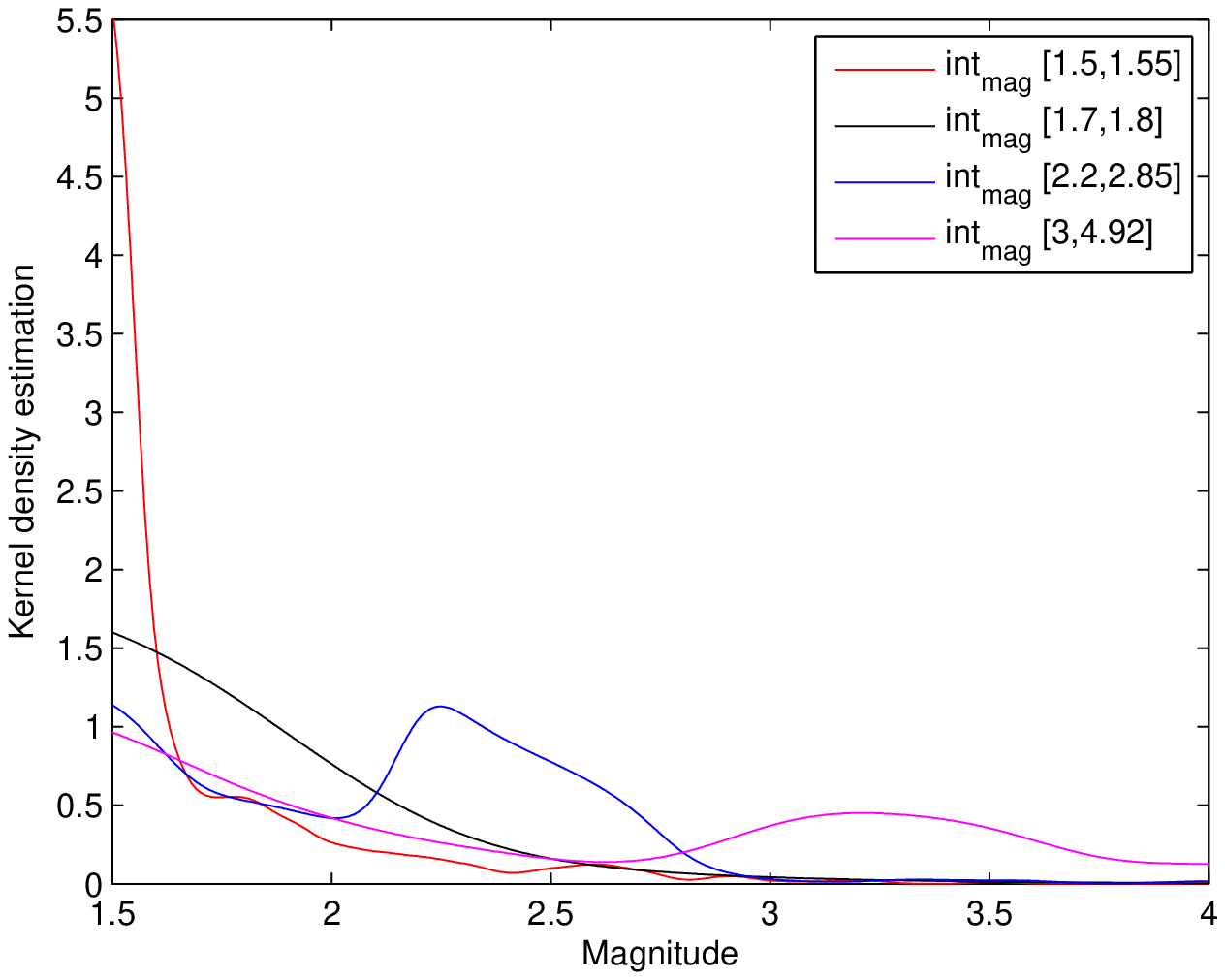}
\includegraphics[width=20pc]{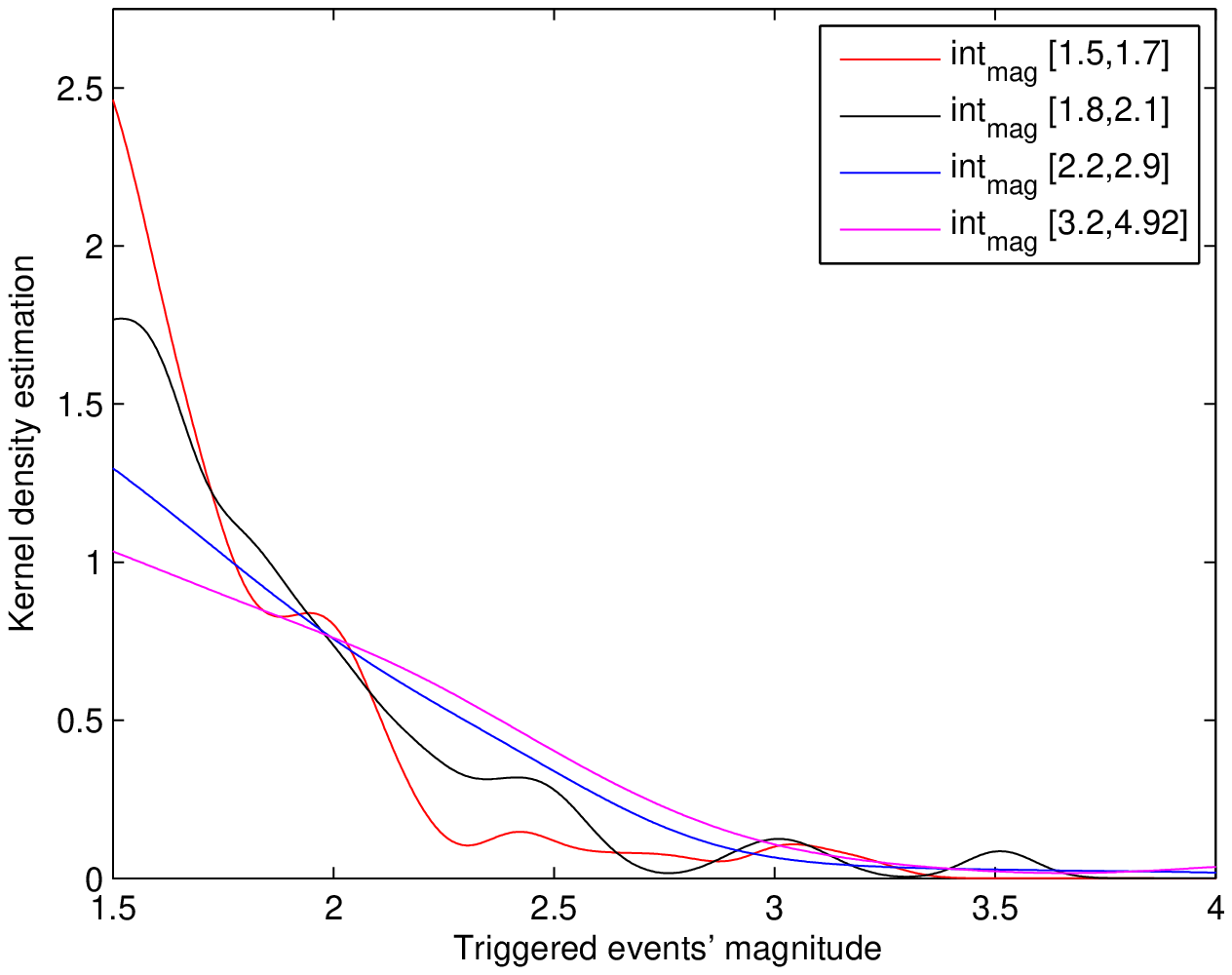}}
   \caption{}
\label{fig:subfig6}
\end{figure}

\begin{figure}
   \centering
\mbox{\includegraphics[width=20pc]{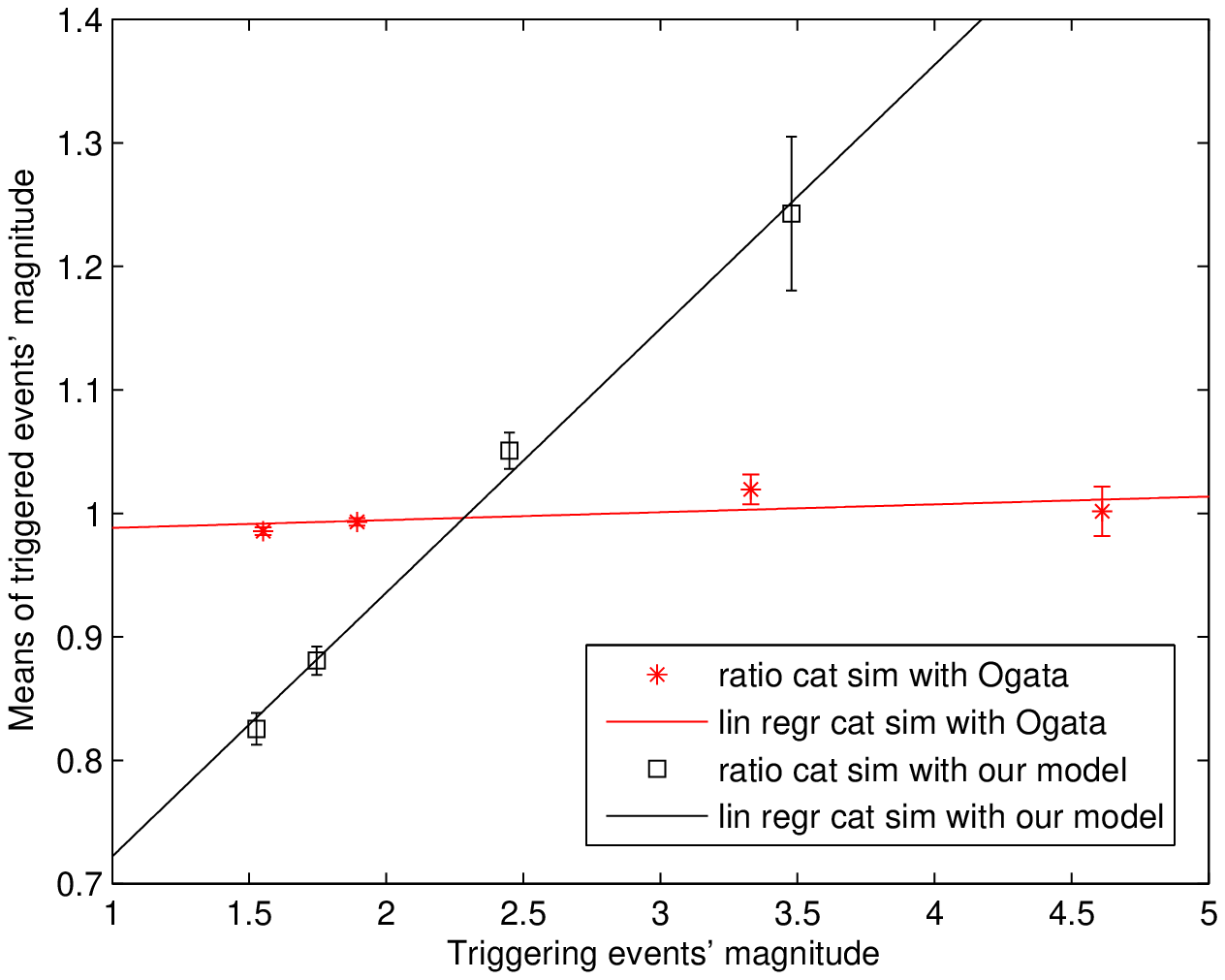}
\includegraphics[width=20pc]{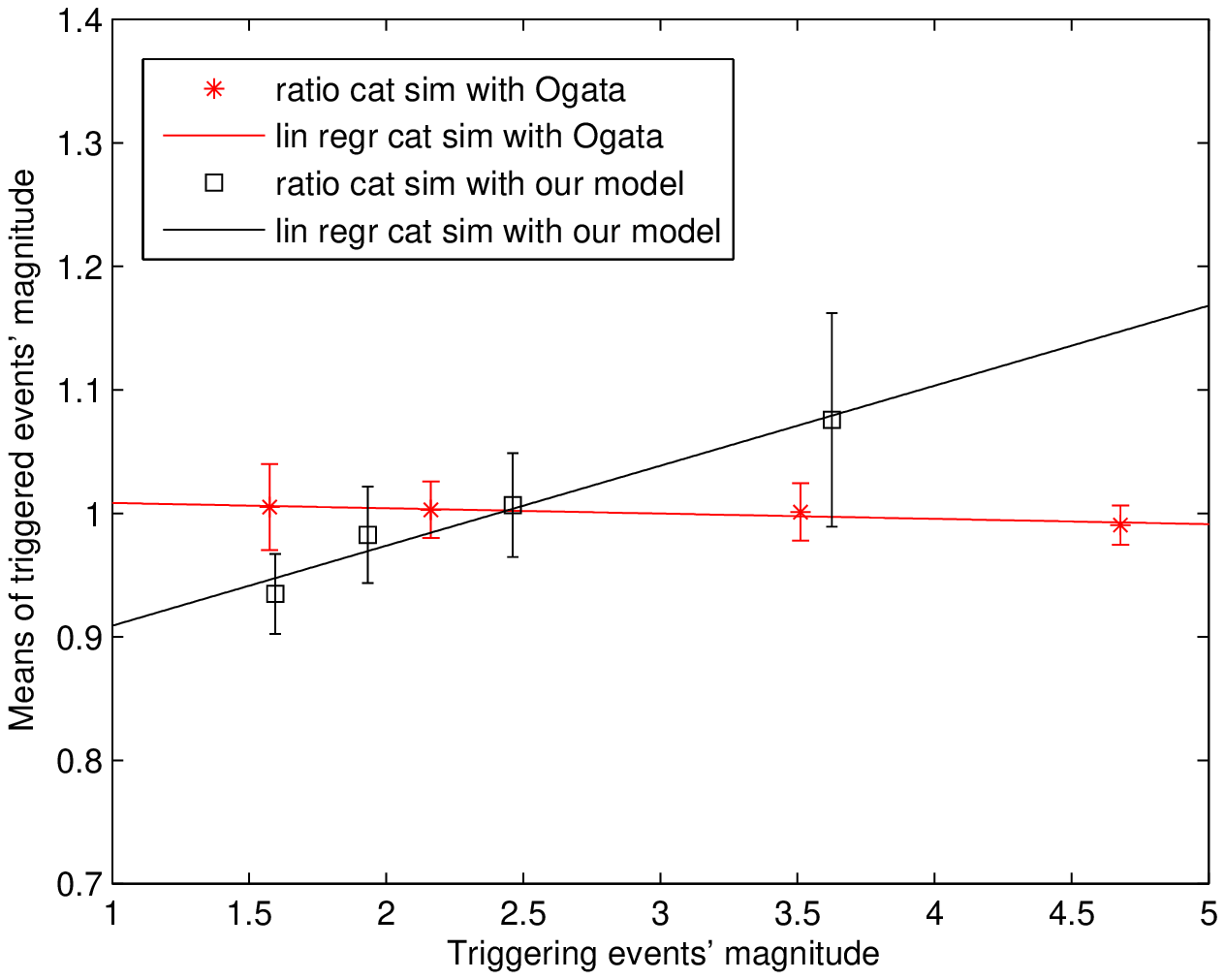}}
\mbox{\includegraphics[width=20pc]{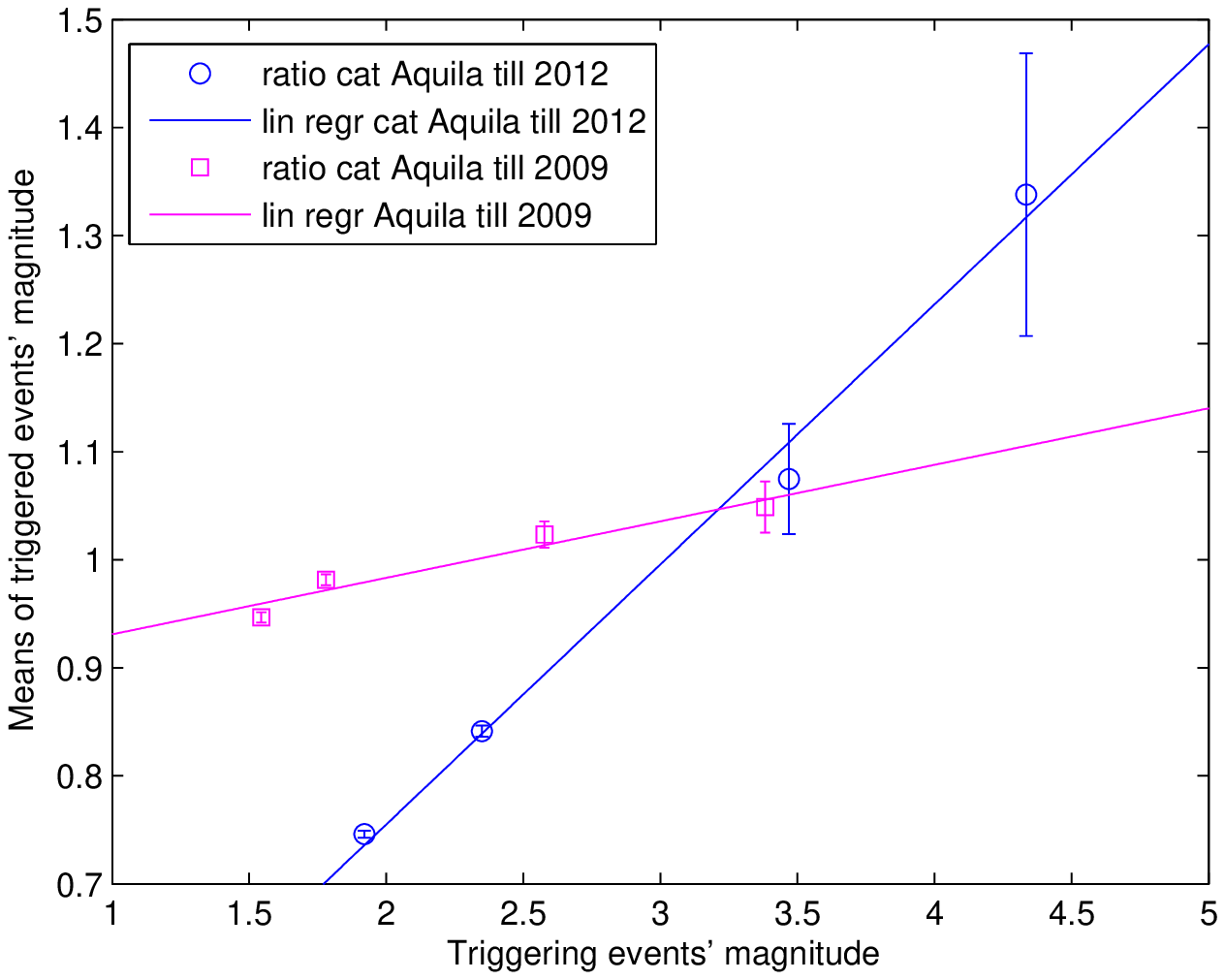}
\includegraphics[width=20pc]{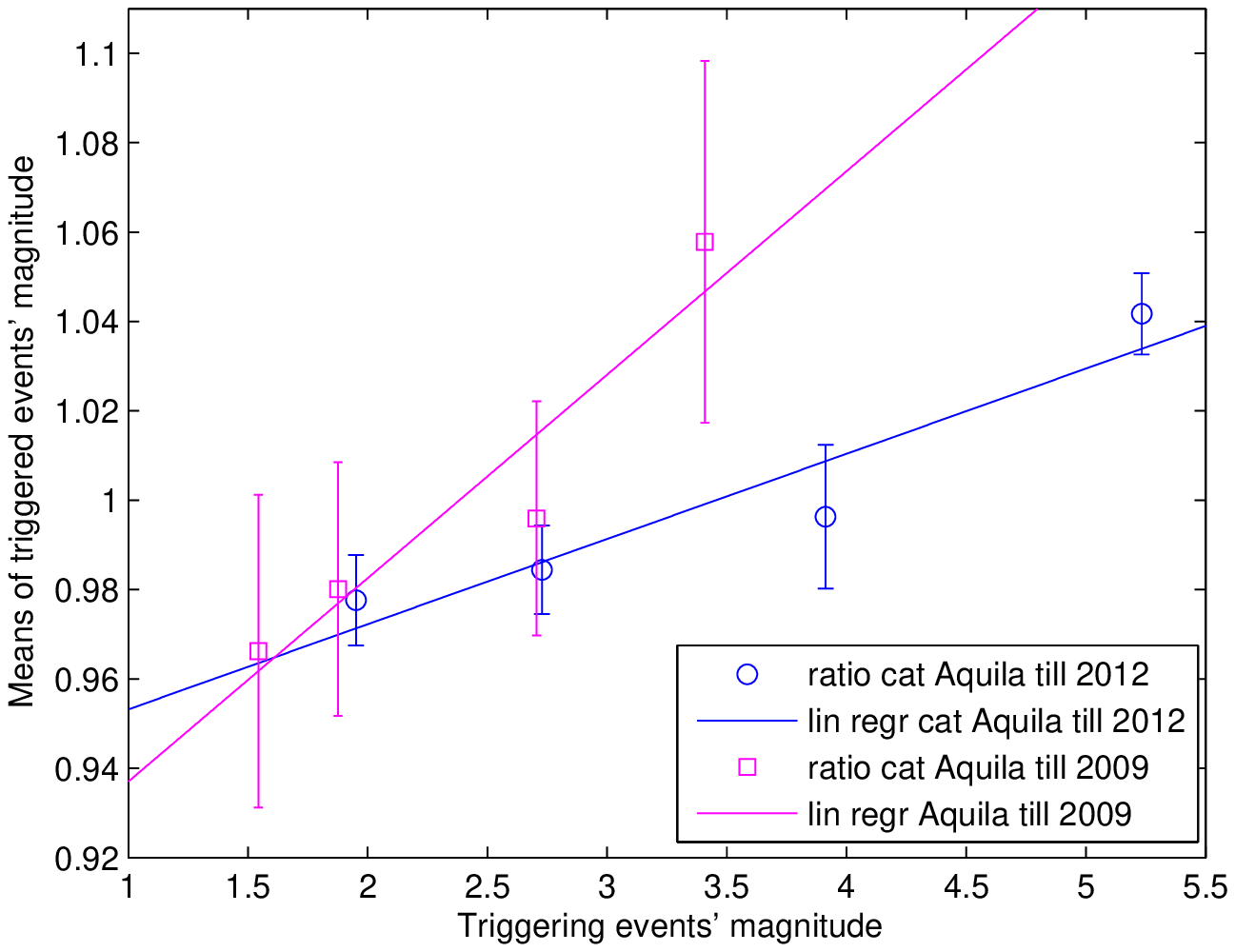}}
   \caption{}
   \label{fig:subfig4}
\end{figure}

\begin{enumerate}
\item Fig.~\ref{fig:subfig2}: Kernel density estimation of triggered events' magnitude concerning the first and the second types of analysis (left and right plots, respectively). At the top, the results concerns the second catalog (L'Aquila till 2012). The considered intervals in which triggering events' magnitudes fall are the following. First analysis, left plot: $[1.8,2.1]$, $[2.2,2.6]$, $[3.3,3.8]$ and $[3.9,5.9]$; second analysis, right plot: $[1.8,2.2]$, $[2.5,3.2]$, $[3.6,4.6]$ and $[4.9,5.9]$ (in both cases, the curves are red, black, blue and magenta, respectively). The ${\delta}^*$ value is equal to one day. The optimal bandwidth value for the Normal kernel density estimation is, respectively for the four intervals considered, equal to: 0.22, 0.33, 0.28, 0.19 in the left plot and 0.25, 0.26, 0.31, 0.27 in the right one.

   In the middle, the results concerns the third catalog (L'Aquila till 5 April 2009). The considered intervals in which triggering events' magnitudes fall are the following. First analysis, left plot: $[1.5,1.6]$, $[1.7,1.9]$, $[2.4,2.9]$ and $[3.1,4.1]$; second analysis, right plot: $[1.5,1.6]$, $[1.8,2]$, $[2.5,3.1]$ and $[3.1,4.1]$ (in both cases, the curves are red, black, blue and magenta, respectively). The ${\delta}^*$ value is equal to four days. The optimal bandwidth value for the Normal kernel density estimation is, respectively for the four intervals considered, equal to: 0.11, 0.14, 0.11, 0.15 in the left plot and 0.14, 0.16, 0.22, 0.15 in the right one.

   At the bottom, the results concerns the fourth catalog (California). The considered intervals in which triggering events' magnitudes fall are the following. First analysis, left plot: $[2,2.2]$, $[2.3,2.39]$, $[2.8,3.2]$ and $[3.3,5.6]$; second analysis, right plot: $[2,2.25]$, $[3.2,3.5]$, $[3.5,4]$ and $[4.6,5.6]$ (in both cases, the curves are red, black, blue and magenta, respectively). The ${\delta}^*$ value is equal to one day. The optimal bandwidth value for the Normal kernel density estimation is, respectively for the four intervals considered, equal to: 0.11, 0.44, 0.12, 0.25 in the left plot and 0.18, 0.13, 0.1, 1 in the right one.
\item Fig.~\ref{fig:subfig6}: Kernel density estimation of triggered events' magnitude concerning the first and the second types of analysis (left and right plots, respectively). At the top, the results concerns the catalog simulated with the classical Ogata's model. The considered intervals in which triggering events' magnitudes fall are the following. First analysis, left plot: $[1.5,1.6]$, $[1.8,2]$, $[3.1,3.6]$ and $[4.13,5.13]$; second analysis, right plot: $[1.5,1.65]$, $[2,2.4]$, $[3.3,3.8]$ and $[4.23,5.13]$ (in both cases, the curves are red, black, blue and magenta, respectively). The ${\delta}^*$ value is equal to seven days. The optimal bandwidth value for the Normal kernel density estimation is, respectively for the four intervals considered, equal to: 0.14, 0.09, 0.08, 0.11 in the left plot and 0.12, 0.11, 0.1, 0.1 in the right one.

   At the bottom, the results concerns the catalog simulated with our conditional model. The considered intervals in which triggering events' magnitudes fall are the following. First analysis, left plot: $[1.5,1.55]$, $[1.7,1.8]$, $[2.2,2.85]$ and $[3,4.92]$; second analysis, right plot: $[1.5,1.7]$, $[1.8,2.1]$, $[2.2,2.9]$ and $[3.2,4.92]$ (in both cases, the curves are red, black, blue and magenta, respectively). The ${\delta}^*$ value is equal to one day. The optimal bandwidth value for the Normal kernel density estimation is, respectively for the four intervals considered, equal to: 0.05, 0.34, 0.07, 0.18 in the left plot and 0.08, 0.08, 0.26, 0.28 in the right one.
\item Fig.~\ref{fig:subfig4}: Averages of triggered events' magnitudes. At the top, the results concerns the catalog simulated with Ogata's model (red) and the one simulated with our model (black), concerning the first and the second types of analysis (left and right plots, respectively). Regarding the left plot, the percentage means for the four subintervals considered for the above two catalogs are the following. Catalog simulated with Ogata's model: 0.9856, 0.9932, 1.0194, 1.0018 (corresponding to the triggering events' magnitude 1.5506, 1.8931, 3.3302, 4.6113, respectively); catalog simulated with our model: 0.8256, 0.8808, 1.0508, 1.2428 (corresponding to the triggering events' magnitude 1.5268, 1.7467, 2.4491, 3.4779, respectively). Regarding the right plot, the normalized means for the four subintervals considered for the two catalogs are the following. Catalog simulated with Ogata's model: 1.0052, 1.0029, 1.0011, 0.9907 (corresponding to the triggering events' magnitude 1.5742, 2.163, 3.5112, 4.6786, respectively); catalog simulated with our model: 0.9349, 0.9827, 1.0067, 1.0757 (corresponding to the triggering events' magnitude 1.595, 1.9324, 2.4614, 3.6254, respectively).

   At the bottom, the results concerns the second and the third catalog, respectively relative to L'Aquila till 2012 (blue) and L'Aquila till 5 April 2009 (magenta), concerning the first and the second types of analysis (left and right plots, respectively). Regarding the left plot, the percentage means for the four subintervals considered for the above two catalogs are the following. Catalog two: 0.7460, 0.8413, 1.0748, 1.3379 (corresponding to the triggering events' magnitude 1.9199, 2.3491, 3.4687, 4.3342, respectively); catalog three: 0.9466, 0.9814, 1.0233, 1.0487 (corresponding to the triggering events' magnitude 1.5433, 1.7796, 2.5770, 3.3818, respectively). Regarding the right plot, the normalized means for the four subintervals considered for the two catalogs are the following. Catalog two: 0.9776, 0.9844, 0.9963, 1.0417 (corresponding to the triggering events' magnitude 1.9508, 2.7277, 3.9125, 5.2333, respectively); catalog three: 0.9662, 0.9801, 0.9959, 1.0578 (corresponding to the triggering events' magnitude 1.5433, 1.8763, 2.7056, 3.4083, respectively).
   The continuous lines correspond to the results of the linear regression and the semi-amplitude of the error bars are the normalized mean standard errors.
\end{enumerate}

\newpage

\end{document}